\documentclass[aps,reprint, prb, longbibliography, superscriptaddress, showpacs,nobibnotes]{revtex4-1}
\usepackage{graphicx}
\usepackage{bm,amsmath,eqnarray}
\usepackage{dcolumn}
\usepackage{natbib,xcolor}
\usepackage[version=4]{mhchem}
\usepackage{upgreek}
\usepackage{braket}
\usepackage{physics}
\usepackage{color}

\begin{document}
\title{Canted antiferromagnetism in a spin-orbit coupled $\bm{S_{\text{eff}}=3/2}$ triangular-lattice magnet \ce{DyAuGe}}

\author{Takashi Kurumaji}
\email{Corresponding author: kurumaji@caltech.edu}
\affiliation{Department of Advanced Materials Science, University of Tokyo, Kashiwa, 277-8561, Japan}
\affiliation{Division of Physics, Mathematics and Astronomy, California Institute of Technology, Pasadena, California 91125, USA}
\author{Masaki Gen}
\affiliation{RIKEN Center for Emergent Matter Science (CEMS), Wako 351-0198, Japan}
\affiliation{Institute for Solid State Physics, University of Tokyo, Kashiwa 277-8581, Japan}
\author{Shunsuke Kitou}
\affiliation{Department of Advanced Materials Science, University of Tokyo, Kashiwa, 277-8561, Japan}
\author{Kazuhiko Ikeuchi}
\affiliation{Institute of Materials Structure Science, High Energy Accelerator Research Organization, Tsukuba, Ibaraki, 305-0801, Japan}
\affiliation{Materials and Life Science Experimental Facility, J-PARC Center, Tokai, Ibaraki 319-1195, Japan}
\affiliation{Graduate Institute for Advanced Studies, SOKENDAI, Tsukuba, Ibaraki, 305-0801, Japan}
\author{Hajime Sagayama}
\affiliation{Institute of Materials Structure Science, High Energy Accelerator Research Organization, Tsukuba, Ibaraki, 305-0801, Japan}
\author{Hironori Nakao}
\affiliation{Institute of Materials Structure Science, High Energy Accelerator Research Organization, Tsukuba, Ibaraki, 305-0801, Japan}
\author{Tetsuya R. Yokoo}
\affiliation{Institute of Materials Structure Science, High Energy Accelerator Research Organization, Tsukuba, Ibaraki, 305-0801, Japan}
\affiliation{Materials and Life Science Experimental Facility, J-PARC Center, Tokai, Ibaraki 319-1195, Japan}
\affiliation{Graduate Institute for Advanced Studies, SOKENDAI, Tsukuba, Ibaraki, 305-0801, Japan}
\affiliation{University of Tsukuba, Tsukuba, Ibaraki 305-8573, Japan}
\author{Taka-hisa Arima}
\affiliation{Department of Advanced Materials Science, University of Tokyo, Kashiwa, 277-8561, Japan}
\affiliation{RIKEN Center for Emergent Matter Science (CEMS), Wako 351-0198, Japan}

\date{\today}
\begin{abstract}
The exploration of nontrivial magnetic states induced by strong spin-orbit interaction is a central topic of frustrated magnetism.
Numerous studies have been conducted on rare-earth-based magnets and 4d/5d transition metal compounds.
These are mostly described by an effective spin $S_{\text{eff}}=1/2$ for the Kramers doublet of the lowest crystal-electric-field levels.
The variety of magnetic orderings can be greatly enhanced when magnetic dipolar moments intertwined with multipolar degrees of freedom, which are described by higher-rank tensors and often require the magnetic ions to have $S_{\text{eff}}>1/2$. 
Here, using synchrotron x-ray diffraction near the Dy $L_3$ edge, we unveil a canted antiferromagnetic ground state arising from a quasi-quartet ($S_{\text{eff}}=3/2$) of 4f electrons in a triangular-lattice (TL) rare-earth intermetallics DyAuGe.
The magnetic moment and electric-quadrupole moment are closely interlocked and a noncollinear magnetic-dipole alignment is induced by antiferroic electric-quadrupole (AFQ) ordering in the TL layers.
The AFQ order is suppressed by an in-plane magnetic field, leading to the metamagnetic transition to a collinear up-up-down magnetic state.
These findings offer insights into the emergence of nontrivial magnetic states in frustrated TL systems with $S_{\text{eff}}>1/2$.
\end{abstract}

\maketitle

\section{Introduction}
Magnetic frustration arises when competing interactions cannot be simultaneously satisfied.
It has been extensively studied in the context of geometrical frustration \cite{ramirez1994strongly} and competing nearest and next-nearest interactions \cite{chandra1990ising}.
Recently, introduction of strong spin-orbit coupling into the interplay with magnetic frustration has been increasingly recognized as a source of exotic magnetic states \cite{witczak2014correlated,takayama2021spin}.
The most prominent example is bond-dependent anisotropic interactions based on an effective spin $S_{\text{eff}}=1/2$ model \cite{jackeli2009mott}, which is found to be relevant to the emergence of a spin-liquid state \cite{kitaev2006anyons,takagi2019concept} as well as unconventional magnetic structures in a broad family of transition-metal \cite{kim2009phase,sagayama2013determination,lu2017magnetism} and rare-earth \cite{li2015rare,li2017kitaev,rau2019frustrated,jang2019antiferromagnetic,clark2019two,li2020partial} compounds.

Another intriguing aspect of spin-orbit coupling that has been less explored in the context of magnetic frustration is the role of multipolar degrees of freedom in intertwining with the spin (magnetic dipole) degree of freedom \cite{santini1999magnetism}.
Multipole moments are allowed in magnetic ions with an angular momentum larger than $1/2$, and have been shown to sometimes give rise to pure higher-order multipolar order in non-frustrated cases of rare earth and actinide compounds \cite{morin1990quadrupolar,kuramoto2009multipole,chen2008spin}.
We also note that such higher-rank multipole degrees of freedom have long been recognized to play an important role also in the physics of cold atoms \cite{wu2003exact,wu2006hidden}.
In the context of frustration, there have been theoretical discussions on novel multipoar liquid or spin-ice phases \cite{lee2012generic,huang2014quantum,li2017symmetry}, multiple-$q$ states involving hybridized dipole and multipole moments \cite{walker1994triple,liu2016semiclassical,ishitobi2023triple}, and flavor-wave excitations \cite{joshi1999elementary,lauchli2006quadrupolar,tsunetsugu2006spin}

The triangular lattice (TL) is a representative playground for the study of the interplay between magnetic frustration and multipole moments. 
To date a number of theoretical models have been proposed on the TL that hosts multipolar degrees of freedom and interactions.
For example, they can be categorized into models with effective spin $S_{\text{eff}}=1/2$ \cite{li2016hidden,liu2018selective,khaliullin2021exchange} and $S_{\text{eff}}=1$ \cite{tsunetsugu2006spin,lauchli2006quadrupolar,amari2022c,zhang2023cp2}, depending on the ground-state multiplet degeneracy.
On the experimental side, although various rare-earth-based TL magnets have been investigated, the majority of the cases are characterized by $S_{\text{eff}}=1/2$ \cite{li2015rare,baenitz2018naybs,shin2020magnetic,ochiai2021field,cong2023magnetic,uzoh2023influence,chamorro2023magnetic} and the observed orderings are a collinear-stripe \cite{xing2021stripe,kulbakov2021stripe} type structures of magnetic dipole moments.
Signatures of multipolar degrees of freedom are discussed in limited materials, i.e., \ce{NiGa2S4} ($S_{\text{eff}}=1$) \cite{nakatsuji2005spin}, \ce{UPd3} ($S_{\text{eff}}=1$) \cite{walker2006determination}, \ce{FeI2} ($S_{\text{eff}}=1$) \cite{bai2021hybridized}, and \ce{TmMgGaO4} (non-Kramers $S_{\text{eff}}=1/2$) \cite{shen2019intertwined}.
Beyond those systems, in principle, a quasi-quartet ($S_{\text{eff}}=3/2$) can also be an ideal platform for studying the interplay between dipolar and mutlipolar interactions \cite{sivardiere1972dipolar,chen2024multi}.

\ce{DyAuGe} is a member of rare-earth-based polar semimetals $R$\ce{AuGe} belonging to the space group $P6_3mc$ (No. 186) \cite{rossi1992ternary}.
The crystal structure is of LiGaGe-type, featuring a TL of Dy atoms coordinated by Au and Ge atoms.
Previous investigations on polycrystalline samples reported antiferromagnetic (AFM) transitions \cite{penc1999magnetic} along with commensurate and incommensurate magnetic modulations in the TL plane \cite{gibson1998investigation}, indicating the presence of magnetic frustration.
Recent efforts by the group of the authors have resulted in the growth of single crystals of $R$\ce{AuGe} \cite{kurumaji2023single}, shedding light on the anisotropic magnetic properties affected by the crystal-electric-field (CEF).
Relatively large magnetoelastic coupling in \ce{DyAuGe} has been observed, which is a potential outcome of the lattice-quadrupole interaction.

In this study, we have revealed the quasi-quartet nature of a Dy$^{3+}$ ion as the lowest CEF multiplets, carrying both magnetic-dipole and electric-quadrupole degrees of freedom by virtue of the $S_{\text{eff}}=3/2$ nature with an energy splitting due to easy-plane anisotropy.
Through a synchrotron x-ray diffraction experiment, we have found that the observed diffraction pattern of the magnetic ground state is commensurate to the lattice but cannot be reconciled with the collinear 1-up/1-down stripe state.
This is in contrast to the case of an isostructural compound \ce{HoAuGe}, where the collinear magnetism was reported \cite{gibson2001crystal} and was confirmed by our resonant x-ray diffraction study.
We have found that the quartet plays an essential role in forming a noncollinear magnetic order accompanied by a symmetry-breaking lattice deformation, which is suggestive of antiferroquadrupole (AFQ) order of 4f electrons in Dy.
Our finding on the present material provides an insight into nontrivial magnetic states in spin-orbit-coupled TL systems enriched by the multipole degrees of freedom.

\section{Results}
\subsection{Quasi quartet on Dy$^{3+}$ in DyAuGe}
\begin{figure*}[t]
	\includegraphics[width = 0.9\textwidth]{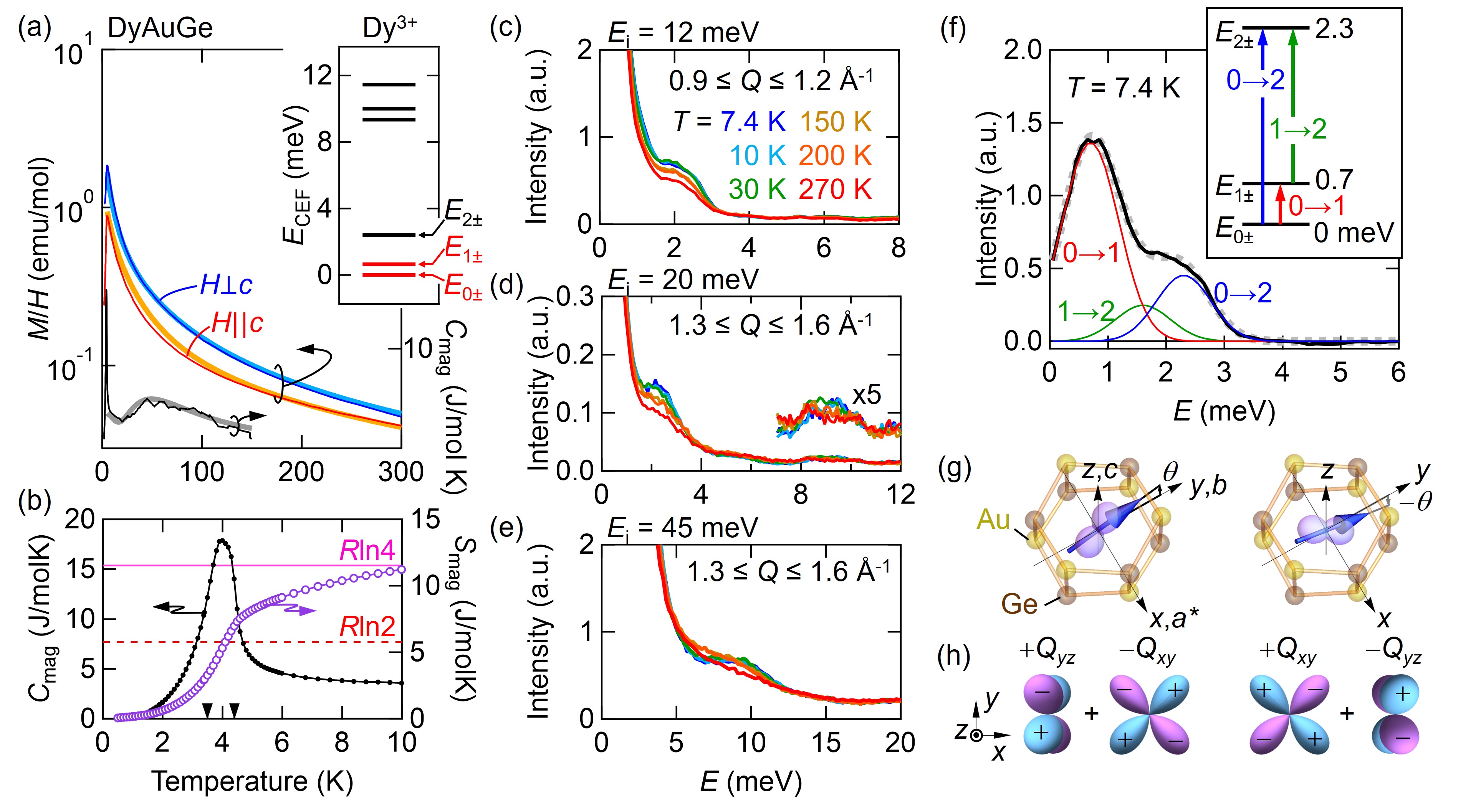}
	\caption{\label{fig1}\textbf{Quasi-quartet with quadrupole moments in \ce{DyAuGe}}
 (a) Temperature dependence of magnetic susceptibility ($M/H$) of a single crystal \ce{DyAuGe} for $H\parallel c$ (red) and $H\perp c$ (blue) at $\mu_0H=0.1$ T and magnetic part of specific heat ($C_{\text{mag}}$, black).
 Thick orange, cyan, and gray curves are fitting by the CEF theory.
 Inset represents the CEF energy levels, calculated with the CEF parameters obtained by the fitting.
 All levels are doubly degenerate.
 The lowest red bars at $E_{0\pm}$ and the second lowest one at $E_{1\pm}$ form the quasi-quartet.
 (b) Temperature dependence of $C_{\text{mag}}$ and magnetic entropy ($S_{\text{mag}}$).
 Black triangles are magnetic transition temperatures ($T_{\text{N1}}$, $T_{\text{N2}}$) determined by the x-ray diffraction study.
 Horizontal red (pink) line corresponds to $S_{\text{mag}}=R\text{ln}2$ ($R\text{ln}4$).
 (c)-(e) INS spectra as a function of energy transfer $E$ for incident neutron energies of (c) $E_{\text{i}}=12$ meV, (d) 20 meV, and (e) 45 meV, measured at several temperatures, in arbitrary units (arb. u.).
 The plot represents intensities integrated within a momentum transfer range between $0.9$ and $1.2$ \AA$^{-1}$ or $1.3$ and $1.6$ \AA$^{-1}$.
 Inset of (d) shows the magnified spectra.
 (f) INS intensity profile, where the elastic scattering contribution is subtracted for clarity.
 Red, blue, and green curves are the individual intensity contributions due to the CEF excitations for $E_{0\pm}\rightarrow E_{1\pm}$, $E_{0\pm}\rightarrow E_{1\pm}$, and $E_{1\pm}\rightarrow E_{2\pm}$, and gray dotted curve is the total intensity.
 Inset shows the schematic CEF energy levels and the excitation processes.
 (g) Simulation of the electric quadrupole moment (purple) at Dy site when the magnetic moment (blue arrow) is canted by $\pm \theta$ from $y$ axis to $\pm z$ axis.
 Cartesian coordinates $y$ and $z$ axes are defined to be parallel to the $b$ and $c$ axes of the crystalline unit cell, respectively.
 $x$ ($a^*$) axis is defined to be perpendicular to both of them.
 The predominant component $Q_{3z^2-r^2}$ is excluded for visibility of other components.
Spherical charge distribution is added as a background.
 Au (Ge) atoms are shown in yellow (brown).
 (h) Schematic illustration of the components of the quadrupole moments induced by the canted magnetic moments in (g).
 }
\end{figure*}

To get insights into the significance of quadrupole moments in Dy$^{3+}$, we estimated the CEF parameters through fitting magnetization and specific heat data.
The results of the fitting are illustrated in Fig. \ref{fig1}(a) (see Sec. VI-E and SI Sec. A for details), showing good agreement with the experiments.
Intriguingly, the lower-lying energy levels, highlighted in the inset, form quasi-quartet composed of two Kramers doublets at $E_{0\pm}$ and $E_{1\pm}$ with a modest energy separation of approximately 0.64 meV, a value comparable to the transition temperature.

The quasi-quartet nature of the lowest CEF levels can also be seen in the specific heat around the magnetic transition temperatures ($T_{\text{N1}}$, $T_{\text{N2}}$).
Figure \ref{fig1}(b) shows the temperature ($T$) dependence of $C_{\text{mag}}$, the magnetic contribution to the specific heat, down to $T=0.5$ K, where the lattice contribution is subtracted from the raw specific heat (see Sec. VI-B and SI Sec. B).
The magnetic entropy $S_{\text{mag}}$ is estimated by integrating $C_{\text{mag}}/T$ with $T$.
With increasing the temperature, $S_{\text{mag}}$ shows a rapid increase at around $T_{\text{N2}}$ due to a broad peak of $C_{\text{mag}}$, and clearly exceeds $R\text{ln}2$ at just above $T_{\text{N1}}$, where $R$ is the gas constant.
Up to $T=10$ K, $S_{\text{mag}}$ approaches to $R\text{ln}4$, corresponding to the entropy release for a quasi-quartet.
To see the contribution from the second excited state at $E_{2\pm}=2.4$ meV, we further analyze the entropy release from given CEF energies and confirm that the quasi-quartet configuration reproduces the $S_{\text{mag}}$ (see detailed analysis in SI Sec. B).

To verify the structure of the CEF levels (inset of Fig. \ref{fig1}(a)), we have performed an inelastic neutron scattering experiment on polycrystalline DyAuGe (see Sec. IV-C and SI Sec. C).
The incident neutron energies $E_{\text{i}}=12$, 20, 45 meV were chosen to suppress the absorption effect.
Figures \ref{fig1}(c)-(e) are the scattering intensity spectra integrated within a certain range of the momentum transfer $|Q|$.
For $E_{\text{i}}=12$ meV (Fig. \ref{fig1}(c)), the spectrum has a broad shoulder below 3 meV which is suppressed with increasing temperature.
Together with the intensity suppression as increasing $|Q|$, this indicates the presence of the low-energy CEF excitations, and the intensity is lowered as the population of the CEF levels changes at higher temperatures.
The excitations at $E< 3$ meV is also evident for the temperature-dependent peak with $E_{\text{i}}=20$ meV (Fig. \ref{fig1}(d)).
The CEF calculation predicts the higher-energy excitations located around $E=9-12$ meV, whose signature is observed as a weak peak for $E_{\text{i}}=20$ meV and a broad peak at $E\sim 10$ meV for $E_{\text{i}}=45$ meV (Fig. \ref{fig1}(e)).

To resolve the low-energy excitations, the scattering intensity for $E_{\text{i}}=12$ meV (Fig. \ref{fig1}(c)) is analyzed with several models (see SI Sec. C for details), where one to three peaks at $0<E< 3$ meV are introduced as the CEF excitations in addition to the intensity for the elastic scattering.
We have found that the three-excitation model gives the best agreement with the observation in the INS experiment.
In this model, we have introduced three intensity peaks, corresponding to two transitions from the ground state, $E_{0\pm}\rightarrow E_{1\pm}$ and $E_{0\pm}\rightarrow E_{2\pm}$, and the transition from the first to second excited state $E_{1\pm}\rightarrow E_{2\pm}$ (see inset of Fig. \ref{fig1}(f)).
The third process is expected to be relevant based on the fact that the first excited state for $E_{1\pm}\sim 8$ K is partially filled at $T=7.4$ K due to the Boltzmann population factor: $e^{-E_{1\pm}/k_{\text{B}}T}$ with $E_{1\pm}\sim k_{\text{B}}T$.
The wave functions given in the CEF calculation also explain the comparable scattering cross section.
We also check that the absence of this third peak gives a self-contradiction to the results of analysis (see SI Sec. C and Fig. S8).

Figure \ref{fig1}(f) shows the intensity profile, where the elastic scattering component is subtracted for clarity.
Within this model, we have identified the low-energy CEF excitation levels at $E_{1\pm}=0.7$ meV and $E_{2\pm}=2.3$ meV, which are in good agreement with the CEF calculation obtained by fitting the thermodynamic properties.
We also confirm that only a slight adjustment of the CEF parameters is sufficient in order to reproduce the CEF energy levels observed in the INS experiment, which does not lead to a significant modification of the wave functions and the induced quadrupole moments at Dy$^{3+}$ ions (SI Sec. C).

Here, we discuss the effect of time-reversal symmetry breaking at temperatures below the magnetic transition temperature.
Within the quasi-quartet, the molecular field polarizes the local magnetic dipole moment, which mixes the lowest doublet with the second lowest one.
This facilitates the emergence of quadrupole moments across the magnetic transition, and the induced quadrupole moment is interlocked with the orientation of the dipole moment (see Sec. IV-E and SI Sec. A).
Figure \ref{fig1}(g) represents the quadrupole moments arising from the canting of the magnetic moment from the $b$ ($y$) axis to the $c$ ($z$) axis, which as we show below is relevant to the magnetic ground state in \ce{DyAuGe}.
Each component of the quadrupole moments, $Q$, is proportional to the expectation value of the quadratic form of total angular momentum $\hat{\mathbf{J}}$.
They are calculated regarding the ground state wave function of a Dy$^{3+}$ ion under the influence of the molecular field polarized along the magnetic dipole moment.

As shown in Fig. \ref{fig1}(g), the quadrupole moment, corresponding to anisotropic charge distribution, extends towards one of the coordinated nonmagnetic atoms.
This is due to the induced $Q_{yz}$ and $Q_{xy}$ components of the quadrupole moment (Fig. \ref{fig1}(h)), which are zero when the dipole moment is orientated along the $y$ axis.
The orientation of these induced quadrupole moments depends on the sign of the canting angle ($\theta$) (see SI Sec. A).
This correlation is due to the dipole-quadrupole intertwining in the quasi-quartet and is expected to affect the magnetic ground state in \ce{DyAuGe}.
We also note that the quasi-quartet carries octupole moments, cubic form of the $\hat{\mathbf{J}}$ \cite{shiina1997magnetic}.
As the octupole moment is a higher-rank tensor than the dipole and quadrupole moments, it is expected to give a minor effect in the current material (see SI Sec. A).

\subsection{Nature of magnetic transitions in DyAuGe}
\begin{figure*}[t]
	\includegraphics[width =  0.8\textwidth]{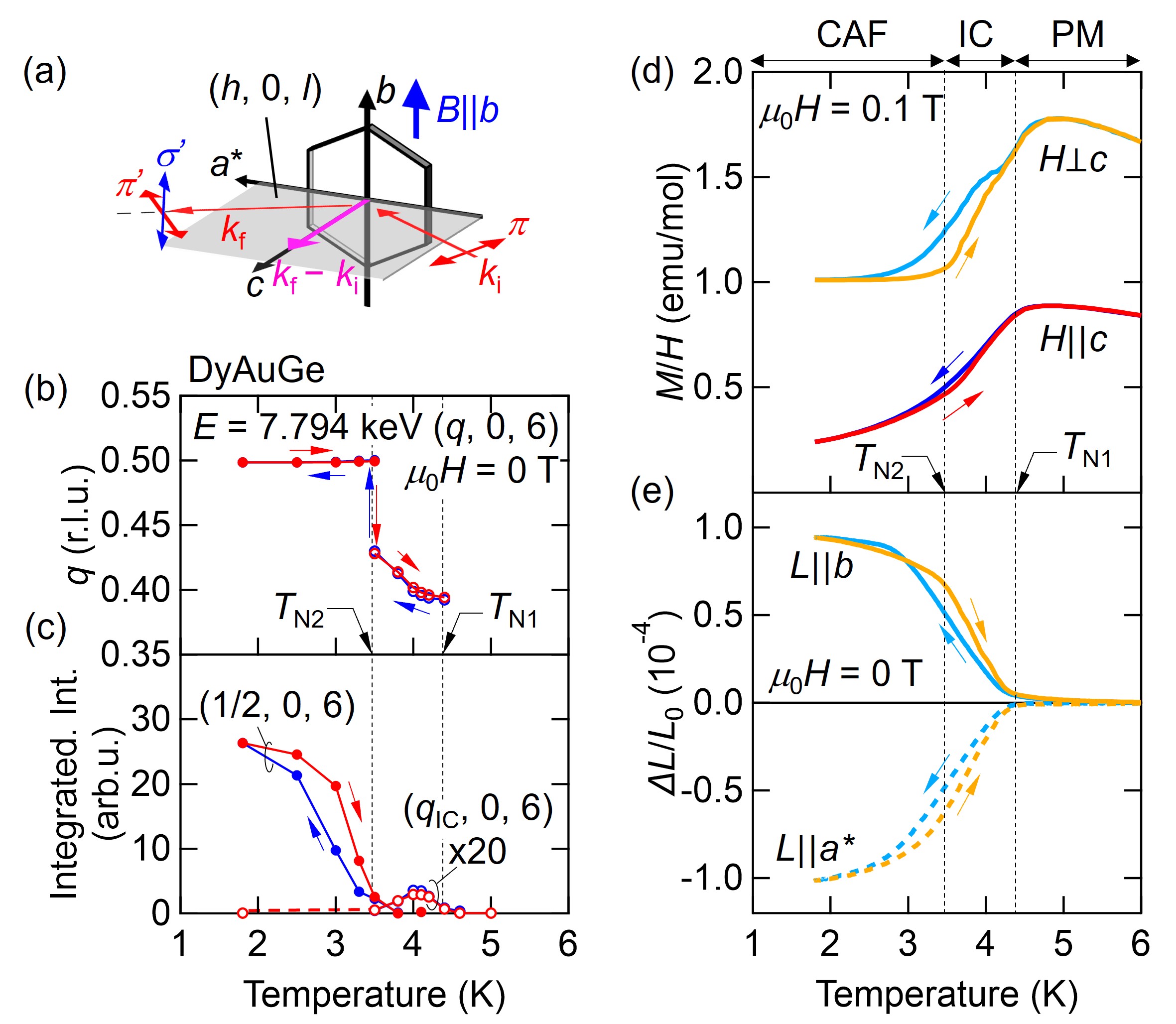}
	\caption{\label{fig2}\textbf{Commensurate and incommensurate states in \ce{DyAuGe}}
 (a) Schematic illustration of x-ray scattering geometry.
 $k_{\text{i}}$ ($k_{\text{f}}$) is the incident (scattered) light.
 The incident polarization of light ($\boldsymbol\epsilon_{\text{i}}$) is set horizontal to the $(h,0,l)$ plane ($\pi$-polarization).
 In the polarization analysis, the polarization of the scattered light ($\boldsymbol\epsilon_{\text{f}}$) is resolved: $\sigma '$ ($\pi '$)-polarization is vertical (horizontal) to the $(h,0,l)$ plane (see Sec. IV-D).
 (b)-(c) Temperature dependence of (b) the modulation wave number $q$ for $(q, 0, 6)$ in the reciprocal lattice unit, and (c) normalized integrated intensity at zero field.
 Red (blue) markers are measured in temperature increase (decrease) process.
 Transition temperatures ($T_{\text{N1}}$, $T_{\text{N2}}$) are denoted.
 (d) Temperature dependence of $M/H$ for $H\parallel c$ and $H\perp c$.
 Arrows indicate the temperature scanning direction.
 Vertical dash lines denote phase boundaries for CAF (commensurate antiferromagnetic), IC (incommensurate), and PM (paramagnetic) states.
 (e) Temperature dependence of thermal expansion along $b$ and $a^*$ axes ($\Delta L/L_0$) near the transition temperatures.
 $L_0$ is defined at $T=6$ K, and $\Delta L=L(T)-L_0$.
 }
\end{figure*}

To see the nature of the magnetic ordering in \ce{DyAuGe}, we perform resonant x-ray diffraction at Dy $L_3$ edge ($E = 7.794$ keV).
The experimental configuration is depicted in Fig. \ref{fig2}(a), where the scattering plane was set to be $(h,0,l)$ (see Sec. IV-D).
According to Refs. \onlinecite{hill1996resonant,hannon1988x}, the resonant x-ray scattering amplitude $f_{\text{res}}$ is expressed as follows,
\begin{equation}\label{eqpol}
    f_{\text{res}}\propto C_0\boldsymbol\epsilon_{\text{f}}^*\cdot \boldsymbol\epsilon_{\text{i}}+\text{i}C_1(\boldsymbol\epsilon_{\text{f}}^*\times \boldsymbol\epsilon_{\text{i}})\cdot \mathbf{m}_{\text{Dy}}+ C_2 \boldsymbol\epsilon_{\text{f}}^{\dagger} \hat{O} \boldsymbol\epsilon_{\text{i}},
\end{equation}
where $C_0$, $C_1$, and $C_2$ are energy-dependent constants, $\mathbf{m}_{\text{Dy}}$ is the magnetic moment at a Dy ion, and $\boldsymbol\epsilon_{\text{i}}$ and $\boldsymbol\epsilon_{\text{f}}$ are the polarization vectors of the incident and scattered x-rays, respectively.
$\hat{O}$ is a tensor proportional to $\mathbf{m}_{\text{Dy}}^{\text{T}}\circ\mathbf{m}_{\text{Dy}}$.
The second term can be a source of an x-ray diffraction through the resonant magnetic scattering.
By analyzing the polarization of the scattered x-ray, the orientation of the magnetic moment $\mathbf{m}_{\text{Dy}}$ is resolved.
The third term is regarding the anisotropic tensor susceptibility (ATS) scattering.
Unless otherwise stated, the scattered x-rays were detected without analyzing polarization of the scattered x-ray and hence include both the $\sigma'$ and $\pi'$ polarizations, which were perpendicular and parallel to the scattering plane, respectively.

We monitor the temperature dependence of the magnetic reflection at $(q, 0, 6)$ in zero field.
Below $T_{\text{N1}}=4.4$ K, we first observe an incommensurate magnetic modulation, consistent with the previous study of powder neutron diffraction \cite{gibson1998investigation}.
The magnetic nature of the incommensurate Bragg peak has been further confirmed by the observation of resonant features in the energy dependence of the scattering intensity (see SI Sec. D).
As the temperature decreases below $T_{\text{N2}}=3.5$ K, the incommensurate magnetic scattering gives way to commensurate Bragg scattering at $(1/2, 0, 6)$, indicating the in-plane doubling of the unit cell.

The successive AFM transitions correlate with magnetic susceptibility and thermal expansion (Figs. \ref{fig2}(d)-(e)).
\ce{DyAuGe} exhibits an easy-plane magnetic anisotropy, as shown in Fig. \ref{fig2}(d).
Below $T_{\text{N1}}$, a hysteresis observed in $M/H$ for $H\perp c$ indicates the breaking of hexagonal symmetry due to the magnetic ordering.
Notably, the $M/H$ for both $H\parallel c$ and $H\perp c$ exhibits a decrease below $T_{\text{N1}}$, signifying the presence of in- and out-of-plane components of magnetic order parameters. 
The six-fold rotational symmetry breaking is further clarified by the opposite signs of thermal expansions along the $a^*$ and $b$ axes, observed below $T_{\text{N1}}$ (Fig. \ref{fig2}(e)).
The magnitude of $\Delta L/L_0$ approaches to $10^{-4}$ at the lowest temperature, which is ten times larger than that in other $R$\ce{AuGe} \cite{kurumaji2023single} and as large as that in \ce{DyB2C2} \cite{yanagisawa2005dilatometric} with quadrupole ordering, suggesting a relevant quadrupole ordering.

\subsection{Energy scan and polarization analysis of x-ray scattering}
\begin{figure*}[t]
	\includegraphics[width =  0.8\textwidth]{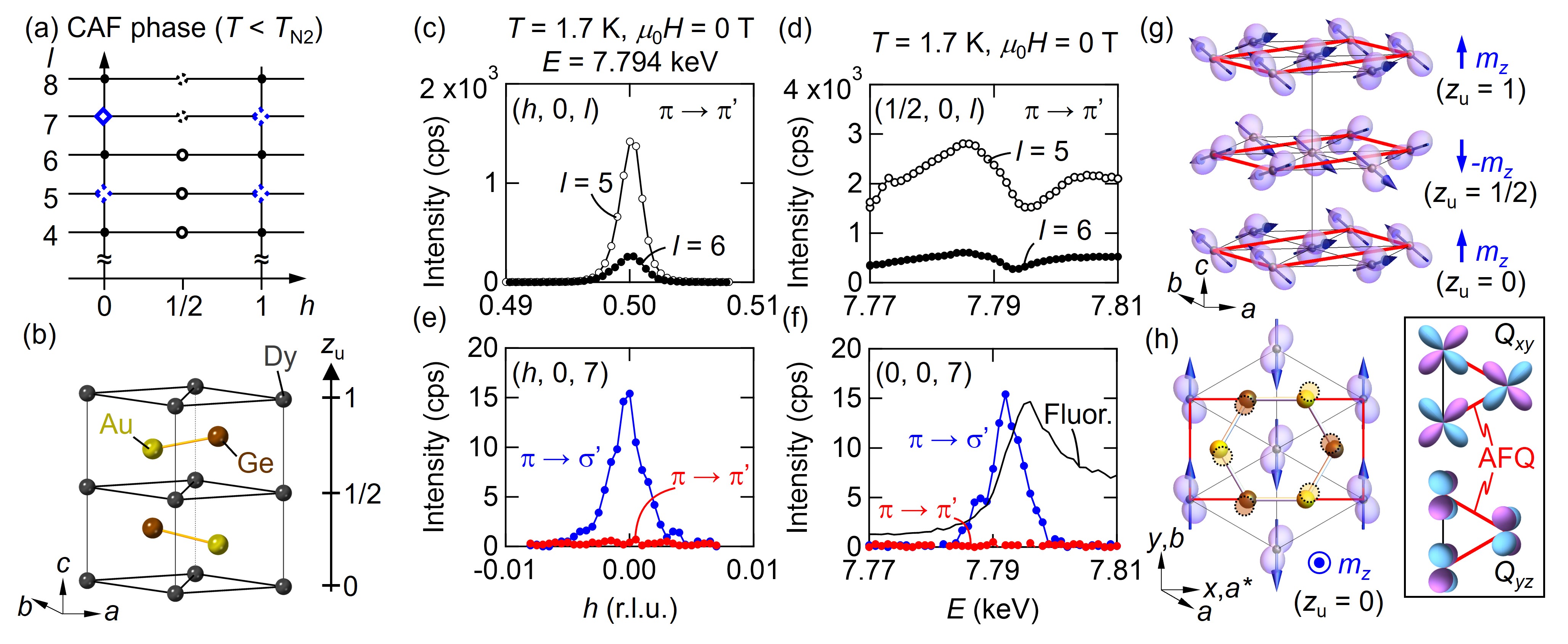}
	\caption{\label{fig3}\textbf{Canted antiferromagnetic state and quadrupole ordering on a triangular-lattice of Dy ions}
 (a) The positions of the Bragg peaks in the $(h,0,l)$ plane at $T<T_{\text{N2}}$.
 $h$ and $l$ are given in reciprocal lattice units.
Black solid circles are the fundamental peaks, open solid (broken) markers represent observed (expected) Bragg peaks.
Circle (diamond) denotes that the energy dependence shows a dip (peak) at the Dy $L_3$ edge.
 (b) Crystal structure of \ce{DyAuGe}.
  $z_{\text{u}}$ is the fractional coordinate along the $c$ axis in a unit cell.
(c)-(d) $h$-scan profile of the scattering intensities for $(h, 0, 6)$ and $(h,0,5)$ measured with the $\pi\rightarrow \pi'$ channel, and the energy dependence of Bragg scattering at $h=1/2$.
(e)-(f) $h$-scan profile of the scattering intensity for $(h,0,7)$ measured in the $\pi\rightarrow \pi'$ and $\pi\rightarrow \sigma'$ channels, and corresponding energy dependence at $h=0$.
Fluorescence excitation spectrum is shown by a thin black curve.
(g) Schematic magnetic structure and quadrupole order for the CAF phase in \ce{DyAuGe}.
Blue vertical arrow on the right side for each $z_{\text{u}}$ layer denotes the out-of-plane magnetic moment ($\pm m_z$) induced by the quadrupole ordering.
(h) Top view of the magnetic structure at $z_{\text{u}}=0$ layer.
Possible shift of Au (Ge) atoms due to the herringbone-type quadrupole alignment is shown by thin yellow (brown) circle.
Inset is a schematic illustration of the AFQ order for the components $Q_{xy}$ and $Q_{yz}$, where AFQ-coupling bonds are highlighted by red lines.
}
\end{figure*}

The Bragg scattering at $(1/2,0,l)$ for $l=$ even has also been observed in a sibling compound \ce{HoAuGe} by our x-ray experiment (see SI Sec. E), the ground state of which is identified as the two-sublattice collinear magnetic state consistent with Ref. \onlinecite{gibson1998investigation}.
In \ce{DyAuGe}, however, we have observed additional features that cannot be accounted for by such a simple magnetic ground state.
Namely, we have observed reflections at $(1/2,0,l)$ for $l =$ odd and also at $(0,0,7)$, as schematically illustrated in Fig. \ref{fig3}(a).
We note that the original lattice has a $c$-glide symmetry.
This would forbid these reflections not only in the paramagnetic state but also in the commensurate AFM (CAF) state if the magnetic moments were merely collinearly aligned in the orthorhombic magnetic unit cell of $\sqrt{3}a \times a\times c$.
The observed reflections indicate that the TL layers at $z_{\text{u}}=0$ and $z_{\text{u}}=1/2$ in the unit cell (Fig. \ref{fig3}(b)) have different magnetic configurations.
We can conclude that the magnetic dipole moments are canted due to the competition of the antiferroic ordering of quadrupole degrees of freedom.

The signature of the AFQ order can be seen in the energy dependence of the Bragg scatterings.
We set the diffractometer in the condition for the Bragg reflections at $(1/2,0,l)$ ($l=$ 5, 6) (Fig. \ref{fig3}(c)) and scan the energy of incident x-ray.
Figure \ref{fig3}(d) shows that the Bragg scattering $(1/2,0,l)$ ($l=$ 5, 6) has a significant intensity even off resonance region and has a dip at the Dy $L_3$ edge due to absorption.
The integrated intensity of these Bragg peaks at an off-resonant condition ($E=7.77$ keV) is only $10^{-2}-10^{-3}$ times smaller than those of the fundamental peaks at $(0,0,6)$ and $(1,0,6)$, in contrast to the fact that the typical magnitude ratio of magnetic scatterings is less than $10^{-4}$.
These indicate that the scattering intensities for $(1/2,0,l)$ ($l=$ 5, 6) are due predominantly to Thomson scattering (the first term in Eq. (\ref{eqpol})) in nature, and resonant feature expected from the magnetic scattering of magnetic moments of Dy is masked due to its relatively weak scattering intensity.
Such a large charge-related scattering is attributed to the displacement of atomic position for elements Au and Ge coupled to the AFQ ordering at Dy site.
We note that the two-sublattice collinear magnetic structure in a TL cannot induce AFQ configuration allowing $(1/2,0,l)$ reflections.

The scattering at $(0,0,7)$ is, in contrast, observed only in the $\pi \rightarrow \sigma'$ channel (Fig. \ref{fig3}(e)), showing a resonant enhancement at Dy $L_3$ edge (Fig. \ref{fig3}(f)).
This is suggestive of the magnetic scattering due to the second term in Eq. (\ref{eqpol}), and indicates that the dipole moments of Dy at each TL layer are canted to have a net magnetic moment ($\mathbf{m}(z_{\text{u}})$) in a certain direction and the orientations at $z_{\text{u}}=0$ and $z_{\text{u}}=1/2$ (Fig. \ref{fig3}(b)) are opposite to each other.
In the current scattering geometry, the scattering in the $\pi \rightarrow \sigma'$ channel is sensitive to the component of the $\mathbf{m}$ lying in the scattering plane, because $\boldsymbol\epsilon_{\sigma '}^*\times \boldsymbol\epsilon_{\pi}$ is parallel to the $a^*c^*$ plane.
Although only the observation of the $(0,0,7)$ peak is not sufficient to resolve the $\mathbf{m}$-component along $a^*$ and $c^*$ axes, dominant $\mathbf{m}\parallel c^*(z)$ component is likely.
The temperature dependence of $M/T$ for $H\parallel c$ (Fig. \ref{fig2}(d)) supports the presence of the out-of-plane antiferromagnetic moments.
We also note that $\mathbf{m}$ with canting towards $\pm a^*$ is compatible to the magnetic point group $2m'm'$ allowing the spontaneous magnetization along the $c$ axis, which is inconsistent with the magnetization measurement (Fig. \ref{fig2}(d)).
Additionally, the $a^*$ component of $\mathbf{m}$ is hard to be reconciled with the absence of a $\pi\rightarrow \pi'$ scattering, which is expected to be observed in the presence of the $C_3$-domains with $\mathbf{q}=(0,-1/2,0)$ and $(1/2,-1/2,0)$.
As another possible mechanism of the resonant scattering for $(0,0,7)$, ATS scattering (third term in Eq. (\ref{eqpol})) induced by the quadrupole moments is also considered \cite{hirota2000direct,okuyama2005quadrupolar}.
Decomposition of the magnetic and ATS scattering is a subject of future study, which is expected to be done by analyzing the interference of E1 and E2 processes \cite{matsumura2005d}.

Figure \ref{fig3}(g) illustrates the schematic configuration of Dy magnetic moments and coupled quadrupole moments, deduced from the observations of the x-ray diffraction.
The magnetic order is close to the two-sublattice collinear state with the magnetic moments in the $\pm b$ direction, while they are canted towards the $c$ axis. 
Canted direction of Dy moments on the layer at $z_{\text{u}}=1/2$ is opposite to that on the layer at $z_{\text{u}} = 0$, being consistent with the observation of the resonant scattering at $(0,0,7)$.
Dictated by the quasi-quartet nature of the lower-lying CEF states, canting of the magnetic moment is concomitant with the anisotropic electronic distribution i.e., a quadrupole moment illustrated by a purple dumbbell.
As shown in Fig. \ref{fig1}(h), the canting of magnetic moment from $y$ ($b$) to $z$ ($c$) axis induces the quadrupole moments, $Q_{xy}$ and $Q_{yz}$, antisymmetric with respect to the canting angle.
Figure \ref{fig3}(h) shows the top view of the quadrupole ordering, which is reminiscent of the herringbone-type ordering as frequently observed in the ordered state of diatomic molecules adsorbed on a graphite substrate \cite{mouritsen1982fluctuation}, where the staggered alignment of quadrupole moments, the atomic bond axis, is stabilized by a nearest-neighbor AFQ interaction.

Possible displacements of Au and Ge atoms within the AFM order are also depicted in Fig. \ref{fig3}(h), which is expected to be induced by the quadrupole moments \cite{adachi2002ordered}, i.e., positively and negatively charged atoms are shifted closer to and away from the lobe of electronic distribution of Dy quadrupole moment, respectively, to gain electrostatic potential.
Since Au and Ge atoms are located between neighboring Dy sites, the atomic shift may be a source of the AFQ interaction between each component of the quadrupole moment (inset of Fig. \ref{fig3}(h)) \cite{nagao2004cooperative,morin1990quadrupolar}.
This may explain the large lattice contribution to the Bragg scattering at $(1/2,0,l)$ (Figs. \ref{fig3}(b)-(e)).
Identification of the dominant AFQ interaction channel is desired to understand the mechanism of the phase transition.

Dzyaloshinskii-Moriya (DM) interaction \cite{dzyaloshinsky1958thermodynamic,moriya1960anisotropic} exists in \ce{DyAuGe} because the polar crystal structure breaks the inversion symmetry.
However, this is not relevant to the proposed canted magnetic structure because the gain of the DM interaction at one Dy-Dy bond in a TL layer is cancelled by the DM interaction energy of the opposite sign at neighboring bond.
The DM interaction along the out-of-plane bonding is also prohibited because of the three-fold rotational symmetry.
Canting mechanism proposed in Ref. \onlinecite{moriya1960theory} can also be excluded because this requires two magnetic sites with different orientations of easy axis. 

Contrast with \ce{HoAuGe} provides additional support for AFQ origin of the noncollinear AFM order in \ce{DyAuGe}.
In \ce{HoAuGe}, the typical two-sublattice collinear state was reported in the previous neutron diffraction study \cite{gibson2001crystal}. 
It is consistent with our resonant x-ray scattering experiment (see SI Sec. E for details).
We have not observed the magnetic scattering other than $(1/2, 0, l)$ for $l=$ even.
The reflection shows a large enhancement at the Ho $L_3$ edge without considerable off-resonant component, indicating the ordering of magnetic dipole moments without significant lattice deformation.
This is consistent with the small thermal expansion at magnetic transition \cite{kurumaji2023single} and the CEF analysis (SI Sec. A), where the quadrupole degrees of freedom is insignificant in \ce{HoAuGe} when the moment is orientated near the $c$ axis.

\subsection{Metamgnetic transitions in an in-plane magnetic field}
\begin{figure*}[t]
	\includegraphics[width =  0.8\textwidth]{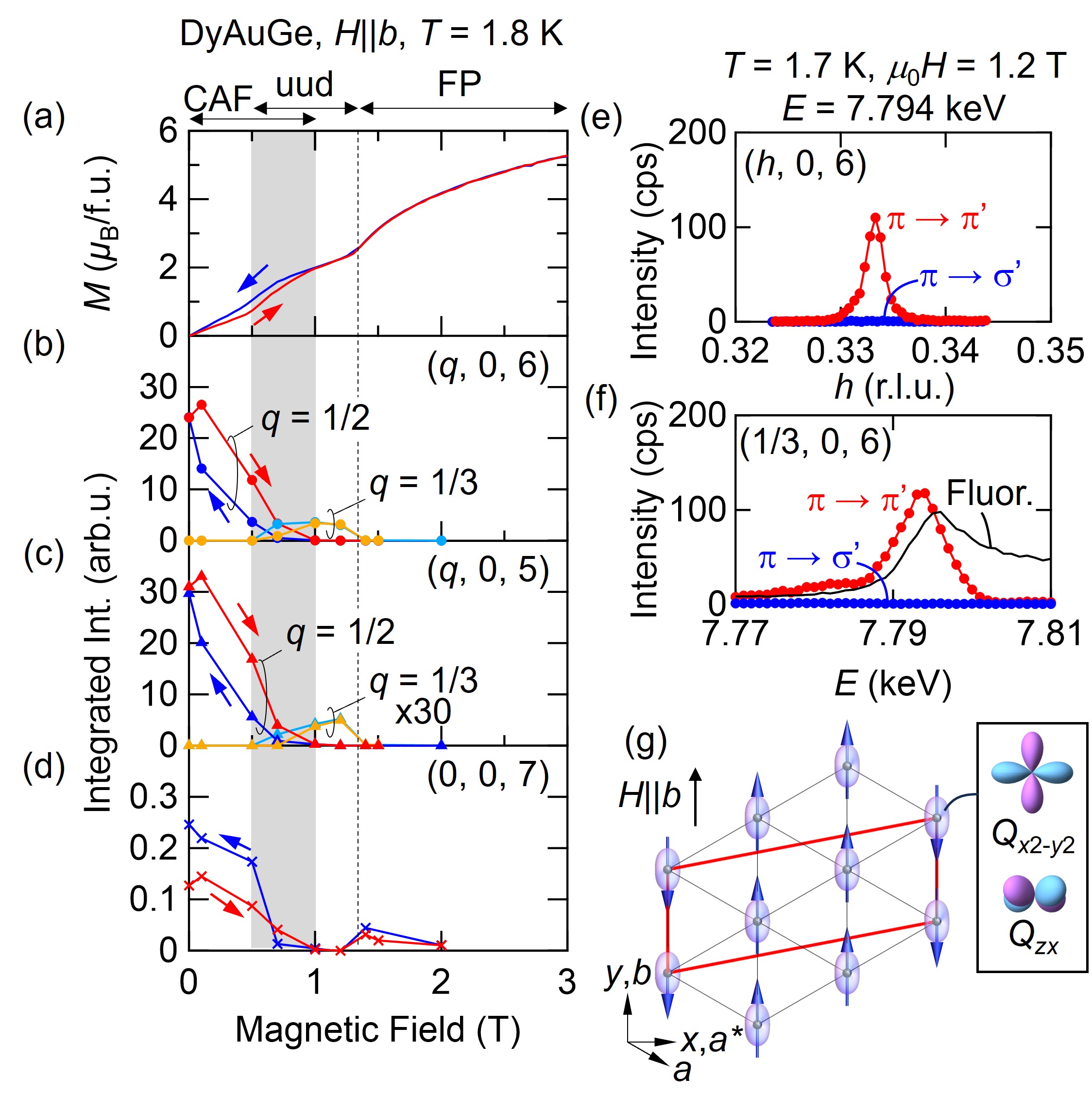}
	\caption{\label{fig4} \textbf{Metamagnetic transition to 1/3-magnetization plateau in \ce{DyAuGe}}
 (a)-(d) Magnetic-field dependence of (a) magnetization ($M$), normalized integrated intensity of Bragg scattering at (b) $(q,0,6)$, (c) $(q,0,5)$, (d) $(0,0,7)$.
 Red (blue) represents the field-increase (decrease) process.
 Gray hatched area denotes a hysteretic region between CAF and uud states.
 Vertical dashed line is guide to eye for a transition field from uud to FP states.
(e)-(f) $h$ scan of the scattering intensity for $(h, 0, 6)$ measured with the $\pi\rightarrow \pi'$ and $\pi\rightarrow \sigma'$ channels, and corresponding energy dependence at $h=1/3$.
Fluorescence is shown by thin black curve, which is scaled to the peak.
 (g) Schematic illustration of the ferrimagnetic structure of \ce{DyAuGe} for $H\parallel b$: magnetic (quadrupole) moment is depicted by blue arrow (purple ellipsoid).
 Red rhombus denotes the magnetic unit cell.
 Inset denotes the components of the ferroquadrupole moments, $-Q_{x^2-y^2}$ and $-Q_{zx}$, induced by the magnetic moment along the $b$ axis.}
\end{figure*}

To clarify the effect of a magnetic field on the AFM/AFQ order, we compare the magnetization and Bragg scattering in an in-plane magnetic field (Fig. \ref{fig4}).
We have found that the AFQ/AFM order in \ce{DyAuGe} is unstable against an in-plane magnetic field and a collinear ferrimagnetic order emerges.
Two-step metamagnetic transitions are observed for $H\parallel b$ (Fig. \ref{fig4}(a)).
The intermediate state emerges near-1/3 magnetization plateau as compared to the moment at higher fields.
The field-dependence of the Bragg peaks shows clear transitions as the peaks at $(1/2,0,l)$ for $l=$ 5, 6 and $(0,0,7)$ disappear, and instead another weaker scattering peak at $q=(1/3,0,6)$ becomes dominant (Figs. \ref{fig4}(b)-(d)).
A component in $q=(1/3,0,5)$ is further weaker than that in $q=(1/3,0,6)$, suggesting that the $c$-glide symmetry is almost resurrected due to the weakening of the influence of AFQ correlation.

Figure \ref{fig4}(e) clarifies the polarization of scattered x-ray for $q=(1/3,0,6)$, which lacks the $\pi \rightarrow \sigma'$ channel contribution.
This scattering shows a resonant peak at Dy $L_3$ edge (Fig. \ref{fig4}(f)), indicating the magnetic origin (the second term in Eq. (\ref{eqpol})).
The $\pi \rightarrow \pi'$ scattering is enhanced by the magnetic moment parallel to the $b$ axis ($\boldsymbol\epsilon^*_{\text{f}}\times \boldsymbol\epsilon_{\text{i}}\parallel b$).
These results indicate that the collinear up-up-down ferrimagnetic state with three sublattices (Fig. \ref{fig4}(g)) is induced as a result of the gain of Zeeman energy.
We note that the magnetic moment along the $b$ axis does not accompany AFQ ordering but a ferroquadrupole (FQ) order with components $-Q_{x^2-y^2}$ and $-Q_{zx}$, as shown in the inset of Fig. \ref{fig4}(g), which is consistent with the weakening of the scattering except $(1/3,0,6)$.

\subsection{Magnetoelastic coupling of quadrupole moments}
\begin{figure*}[t]
	\includegraphics[width = 0.8 \textwidth]{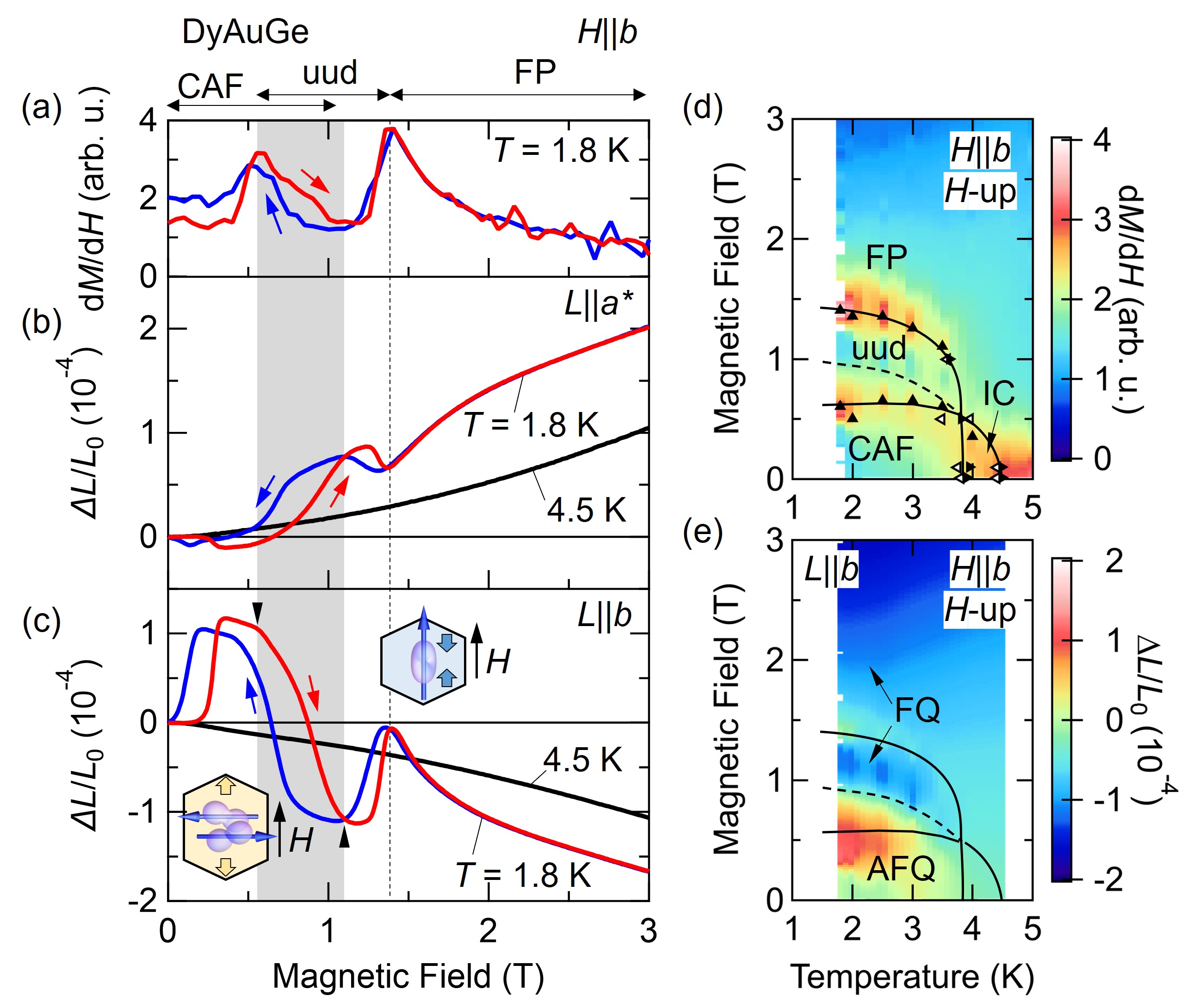}
	\caption{\label{fig5} \textbf{Magnetoelastic coupling of AFQ-FQ transition in \ce{DyAuGe}}
 (a) Field derivative of magnetization (d$M$/d$H$) for $H\parallel b$ at $T=1.8$ K.
  Red (blue) curve is field-increase (decrease) process.
 (b)-(c) Magnetic-field dependence of magnetostriction along (b) $L\parallel b$ and (c) $L\parallel a^*$ $H\parallel b$ at $T=1.8$ K.
 $L_0$ is determined as the value in zero field at the measuring temperature.
 Data at $T=4.5$ K is shown by black curve.
 Gray hatched area is the intermediate region for the transition between CAF and uud states, which is determined by the deflection of $\Delta L/L_0$ along the $b$ axis for the field-increasing scan.
(d) $H-T$ magnetic phase diagram for $H\parallel b$.
Color map of d$M$/d$H$ for $H$-increasing process is overlaid.
Triangle facing right (left) denotes a peak in d$M$/d$T$ in a temperature-increasing (decreasing) process.
Triangle facing up denotes a peak in d$M$/d$H$ in a field-increasing process.
IC: incommensurately modulated magnetic state; CAF: commensurate AFM state; uud: collinear up-up-down type ferrimagnetic state; FP: field-polarized state.
Solid black curve denotes the phase boundary determined by the magnetization measurements.
Dashed line is the upper limit of CAF state suggested in (a)-(c).
(e) Color map of $\Delta L/L_0$ along the $b$ axis for $H\parallel b$ in the $H$-increasing process.
AFQ (FQ) denotes the antiferroquadrupole (ferroquadrupole) order concomitant with the CAF (uud) states.
}
\end{figure*}

To further corroborate the field-induced AFQ-FQ transition, we have measured the lattice distortion as sweeping the in-plane magnetic field.
Figures \ref{fig5}(a)-(c) compare anomalies in the field-derivative of magnetization ($\text{d}M/\text{d}H$) and those in magnetostriction along the $a^*$ and $b$ axes across the metamagnetic transitions.
$\text{d}M/\text{d}H$ shows a broad peak at around 0.5 T with a hysteresis for the field-increasing and decreasing processes.
There is a similar hysteresis in the magnetostriction curve for both $L\parallel a^*$ and $L\parallel b$.
The $L\parallel b$ in field-increase process shows deflections of the slope at fields denoted by triangles in Fig. \ref{fig5}(c), which is identified as a transition region between CAF and uud state.
The lattice elongates along the $b$ axis at low fields of the CAF state, while it shrinks as the field is evolved to induce the uud state to polarize the magnetic moment along the $b$ axis.
This behavior is consistent with the reorientation of the quadrupole moments in the magnetic field as depicted in the inset.
Within this scenario, the AFM moments are oriented to be perpendicular to the $H$ at low fields, which aligns the electron distribution in the horizontal direction and expands the lattice along the $b$ axis.
In the ferrimagnetic and fully polarized states, the FQ moment is elongated along the magnetic moment and induces a lattice shrinkage along the $b$ axis.

Figures \ref{fig5}(d)-(e) show color maps of $\text{d}M/\text{d}H$ and $\Delta L/L_0$ along the $b$ axis in the field-increasing process for $H\parallel b$. 
At $H=0$, the noncollinear AFM order is stabilized due to the dipole-quadrupole coupling.
A magnetic field favors the collinear ferrimagnetic state, which suppresses the AFQ ordering.
As shown in Fig. \ref{fig5}(e), the lattice expansion (red region) is correlated with the AFQ ordering for the CAF phase and the shrinkage (blue region) comes together with the FQ state in the uud and FP phases.
These results confirm the close coupling between the quadrupole moments and the crystal lattice.
We note that a peak is observed in the magnetic-field dependence of $\Delta L/L_0$ along the $b$ axis at the transition field between uud and FP states, suggesting a presence of AFQ correlation at this field.
It is consistent with a small intensity at $(0,0,7)$ right above the second critical field (see Fig. \ref{fig4}(d)), which may suggest a signature of a slight canting of the magnetic moment at the transition between two collinear phases.

\section{Discussion}
Canted antiferromagnetism in \ce{DyAuGe} can be phenomenologically described as follows.
The exchange interaction between magnetic moments is expected to be proportional to $J_{\text{d}}\cos 2\theta$ as a function of the canting angle $\theta$ defined in Fig. \ref{fig1}(g).
$J_{\text{d}}$ ($<0$) is the exchange constant between neighboring dipole moments and favors collinear AFM state ($\theta=0$).
The induced AFQ moments $Q_{xy}$ and $Q_{yz}$ is correlated with $\theta$ as shown in SI Sec. A, which is proportional to $\sin 2\theta$.
Using the AFQ interaction constant $K_{\text{q}} (<0)$, the AFQ interaction energy is proportional to $K_{\text{q}}\sin^2 2\theta$, which is minimized at $\theta=\pi/4$.
The above two terms are competing and a canted configuration is favored when $J_{\text{d}}/2K_{\text{q}}<1$.

The explanation above does not include the effect of magnetic frustration in a TL.
Constructing a microscopic TL model to reproduce the magnetic structure in \ce{DyAuGe} is a possible future study.
Various theories discussed the magnetic ordering in a TL of magnetic ions with strong spin-orbit coupling \cite{li2016hidden,liu2018selective,khaliullin2021exchange,li2016anisotropic,luo2017ground,zhu2018topography,maksimov2019anisotropic,wang2021comprehensive,jackeli2015quantum}.
These studies isolate the lowest doublet from the excited states of CEF levels and develops an effective spin Hamiltonian for $S_{\text{eff}}$-1/2 with anisotropic exchange interactions.
Forms of spin Hamiltonian depend on Kramers or non-Kramers nature of the magnetic ions \cite{liu2018selective}.
For a system with Kramers ions, the collinear magnetic structures are predicted with the modulation vector $\mathbf{q}=(1/2,0,0)$ \cite{li2016anisotropic,luo2017ground,zhu2018topography,maksimov2019anisotropic,wang2021comprehensive}, while noncollinear ordering as observed in \ce{DyAuGe} has not been demonstrated.
In the present system, the quasi-quartet nature of the Dy CEF levels may be beyond the treatment of $S_{\text{eff}}=1/2$.
In order to construct an effective spin Hamiltonian, one may need the formalism based on $S_{\text{eff}}=3/2$ with biquadratic/bicubic forms of exchange interactions, which can lead to interaction terms between quadrupole moments \cite{sivardiere1972dipolar,klevets2018phase}.
We note that the current system is a metal and the RKKY interactions may play a dominant role in the magnetic interactions.
This would also make the anisotropic exchange interactions relatively irrelevant compared to insulating materials.
Consideration of the long-range nature of the exchange and multipolar interactions in the reciprocal space may also be a relevant approach to this open question \cite{hanzawa2019origin,ishitobi2023triple}.

We also note a possible connection through the underlying physics to various systems in condensed matter physics.
Although the direct comparisons between intermetallics and oxides must be made with caution, the noncollinear magnetic ground state in \ce{DyAuGe} is reminiscent of the canted magnetic configuration in double-perovskite magnets with transition metal ions of d$^1$ configuration \cite{lu2017magnetism,hirai2019successive}, where quadrupolar ordering due to a spin-orbit quartet for $J_{\text{eff}}=3/2$ provides a reasonable explanation \cite{chen2010exotic}.
The alkali-earth-like atoms with a nuclear spin of $F$ are known to be described by SU($2F+1$)\cite{taie2010realization,taie20126}.
Some of them are characterized by $F=3/2$, and described by a hidden SO(5) symmetry in a cold atom system \cite{wu2003exact,wu2006hidden}.
Exploring a unified view of the physics in different systems would be helpful to deepen the understanding of complex systems and would facilitate communication between different branches of condensed matter physics.

In contrast to frustrated systems described by $S_{\text{eff}}=1/2$ doublets, magnetism of the quartet system requires a proper treatment of multipole degrees of freedom using higher-rank operators.
The magnetism of a TL with such a quartet has been discussed in the context of a spin-orbital fluctuating $\text{SU}(4)$ Kugel-Khomskii model \cite{li19984,penc2003quantum,keselman2020emergent,yamada20214,jin2022unveiling,zhang2023variational} and multiflavor Mott insulators for $J=3/2$ \cite{chen2024multi}, but rarely explored in real materials.
Our findings in this study is expected to motivate further studies for unexplored magnetic states enriched by a spin-orbit-coupled quartet on a TL system.
We also note that anomalous transport properties in the isostructural $R$AuGe ($R$: rare earth) have recently been reported \cite{ram2023multiple,kurumaji2024metamagnetism}, where a connection to a nodal-line semimetal state is proposed \cite{kurumaji2025electronic}.
These features may potentially provide an unconventional interplay between conduction electrons and multipole moments.

\section{Methods}
\subsection{Single crystal growth}
Single crystals of $R$AuGe ($R$ = Dy, Ho) were grown by using a Au-Ge self-flux as reported in Ref. \onlinecite{kurumaji2023single}.
Starting materials are $R$ ingot (99.9\%), Au wire (99.5\%), and Ge pieces (99.999\%).
They were loaded in a molar ratio of $R$:Au:Ge of 1:2:2 into an alumina crucible \cite{canfield2016use}, which was sealed in an evacuated quartz tube.
The tube was heated to 1100 $^{\circ}$C and held at this temperature for 24 hours before being slowly cooled to 800 $^{\circ}$C in 200 hours.
After being held at 800 $^{\circ}$C for 4 days, the tube was removed from the furnace and centrifuged to remove excess flux.
The obtained crystals had a platelet shape with typical dimensions 1$\times$1$\times$0.5 mm$^{3}$. 
The crystal structure analysis was performed by using a single-crystal x-ray diffractometer on BL02B1 at a synchrotron facility SPring-8, Japan.
We confirm that the crystal structure belonging to $P6_3mc$ \cite{kurumaji2023single} is preserved down to 30 K for both DyAuGe and HoAuGe by the absence of qualitative change of diffraction pattern and intensity.

\subsection{Specific heat, magnetization and magnetostriction measurements}
Specific heat was measured using a commercial system (heat capacity option with helium-3 refrigerator of a Quantum Design PPMS).
To obtain the magnetic part $C_{\text{mag}}$ of $R$AuGe ($R=$ Dy, Ho), the specific heat of LuAuGe was used as a background.
Magnetization was measured with a superconducting quantum interference device magnetometer (Quantum Design MPMS-XL).
Magnetostriction was measured by the fiber-Bragg-grating (FBG) technique using an optical sensing instrument (Hyperion si155, LUNA) in an Oxford Spectromag.
Optical fibers were glued using epoxy (Stycast1266) on the (001) surface of as-grown crystals to simultaneously measure the striction along the $b$ and $a^*$ axes in an in-plane magnetic field \cite{gen2022enhancement}.

\subsection{Inelastic neutron scattering in DyAuGe}
To observe CEF excitations in DyAuGe, we performed inelastic neutron scattering (INS) experiments on a time-of-flight (TOF) spectrometer (POLANO) at the Materials and Life Science Experimental Facility (MLF) of J-PARC in Japan \cite{yokoo2013newly}.
Polycrystalline DyAuGe was prepared by arc melting of stoichiometric elements: Dy ingot (99.9\%) with natural isotopic abundance, Au wire (99.95 \%), are Ge chunks (99.999 \%).
Phase purity was checked by powder x-ray diffraction and the magnetic transitions were confirmed by magnetization measurement.
The sample pieces, weighing a total of 1.6 g, were tiled on an aluminum plate and placed in a GM refrigerator to control the temperature from 7.4 K to 270 K.
The chopper frequency was set at 200 Hz, and the incident energies of $E_{\text{i}}=12$, 20, 45 meV were used to observe CEF excitations.
The intensity integration was typically taken for twelve hours at each temperature.

\subsection{Synchrotron resonant x-ray diffraction}
Single-crystal resonant x-ray scattering experiment was carried out at BL-3A, Photon Factory, KEK, Japan, following the procedure described in Ref. \onlinecite{kurumaji2022anisotropic}.
We used horizontally polarized x-ray near the resonance with Dy $L_3$ absorption edge (7.794 keV).
An as-grown (001) plane of a crystal is attached on an aluminum plate and loaded into a vertical-field superconducting magnet with the $b$ axis parallel to the magnetic field direction.
For the polarization analysis, the 006 reflection of a pyrolytic graphite (PG) crystal was used.
The scattering angle at the PG analyzer was approximately 91$^{\circ}$ for the x-rays near the Dy $L_3$ edge.

\subsection{Crystal electric field theory and quadrupole moments}
To gain insight to the quadrupole degrees of freedom in \ce{DyAuGe} and \ce{HoAuGe}, we analyzed the magnetic susceptibility ($M/H$) and magnetic part of the specific heat ($C_{\text{mag}}$) of these compounds based on the crystal electric field theory.
Fitting of $M/H$ and $C_{\text{mag}}$ was performed by using Mantid \cite{mantid10analysis}.

Crystal electric field (CEF) Hamiltonian, $H_{\text{CEF}}$, for a rare-earth ion at a site with $C_{3v}$ symmetry is given by
\begin{align}
\label{HCEF}
H_{\text{CEF}}=B_{20}\hat{O}_2^0+B_{40}\hat{O}_4^0+B_{60}\hat{O}_6^0\notag\\
+B_{43}\hat{O}_4^3+B_{63}\hat{O}_6^3+B_{66}\hat{O}_6^6, \\
\notag
\end{align} 
where $\hat{O}_n^m$ and $B_{nm}$ are the Stevens operators and parameters, respectively.
$\hat{O}_n^m$ is expressed by the total angular momentum operators, $\hat{J}_x$, $\hat{J}_y$, and $\hat{J}_z$, of $J=15/2$ for Dy$^{3+}$ and $J=8$ for Ho$^{3+}$.
As for Dy$^{3+}$, the 16-fold states for $J=15/2$ are split into eight doublets.
Because of the Kramers nature of a Dy$^{3+}$ ion, all the levels are doubly degenerate.
The temperature dependence of $\chi_{0\parallel c}$, $\chi_{0\perp c}$ without the effect of molecular field, and $C_{\text{mag}}$ is obtained by the formulae (see e.g. Ref. \onlinecite{van2007magnetic}):
\begin{multline}
\chi_{0\parallel c}=\frac{N_{\text{A}}(g_J\mu _{\text{B}})^2}{k_{\text{B}}T}\notag\\
\qquad\qquad\times \sum_{m,n}^{E_m=E_n}\frac{e^{-E_m/k_{\text{B}}T}\mel{m}{\hat{J}_z}{n}\mel{n}{\hat{J}_z}{m}}{Z}, \qquad   (3)\\
\end{multline}
\begin{multline}
\chi_{0\perp c}=\frac{N_{\text{A}}(g_J\mu _{\text{B}})^2}{Z}\notag \\
\times  \sum_{m,n}^{E_m\neq E_n}\mel{m}{\hat{J}_x}{n}\mel{n}{\hat{J}_x}{m}\frac{e^{-E_n/k_{\text{B}}T}-e^{-E_{m}/k_{\text{B}}T}}{E_{m}-E_n}, \quad  (4)\\
\notag
\end{multline}
\begin{multline}
\label{Cmag}
C_{\text{mag}}=\frac{N_{\text{A}}}{k_{\text{B}}T^2}\left[ \left(\frac{1}{Z}\sum_{n} E_n^2e^{-E_n/k_{\text{B}}T} \right)\right. \notag\\
\left.\qquad\qquad\qquad - \left( \frac{1}{Z}\sum_{n} E_n e^{-E_n/k_{\text{B}}T}\right)^2 \right],\qquad\qquad      (5)\\
\notag
\end{multline}
where $N_{\text{A}}$ is Avogadro constant, $k_{\text{B}}$ is Boltzmann constant, $\mu_{\text{B}}$ is Bohr magneton, $g_{J}$ is Land\'{e}'s $g$ factor, $E_{n}$ is the energy of the $n$-th state, $\ket{n}$.
$Z=\sum _{n}e^{-E_n/k_{\text{B}}T}$ is the partition function.
The molecular field is treated by introducing an effective exchange interaction $\lambda$.
The magnetic susceptibilities ($\chi_{\parallel c}$, $\chi_{\perp c}$) are obtained by
\begin{equation}
    \chi=\frac{\chi_0(T)}{1-\lambda\chi_0(T)}.
\end{equation}

In this study, we considered Stevens parameters $B_{43}$, $B_{63}$, and $B_{66}$ in addition to the parameters, $B_{20}$, $B_{40}$, and $B_{60}$, used in the previous paper \cite{kurumaji2023single}.
The results of the fitting for \ce{DyAuGe} are shown in Fig. \ref{fig1}(a) (see SI Sec. A for detailed parameters).
The calculated CEF energies provide a quasi-quartet for the lowest four levels as shown in the inset of Fig. \ref{fig1}(a) with an energy separation of around 8 K between the lowest and second lowest Kramers doublets, which is consistent with the observation in the INS experiment.
The energy splitting is comparable to the transition temperature ($\sim 4$ K).
The molecular field by the exchange interaction is expected to mutually hybridize the quasi-quartet to induce the quadrupole moment.

To evaluate the quadrupole moments at Dy$^{3+}$ ions coupled to the magnetic moment, we projected the CEF Hamiltonian into an effective $4\times 4$ Hamiltonian spanned by the quasi-quartet ($\ket{0\pm}$, $\ket{1\pm}$), which diagonalizes the CEF Hamiltonian.
\begin{equation}
    H_{\text{quartet}}=P^{-1}H_{\text{CEF}}P,
\end{equation}
where $P$ is the projection operator, and $P^{-1}H_{\text{CEF}}P=\text{diag}\{E_{0},E_{0},E_{1},E_{1}\}$.
We calculated the ground state under the effect of a molecular field:
\begin{equation}
    H_{\text{mol}}(\varphi)=J_{\text{ex}}\langle J_{\mathbf{n}}\rangle \mathbf{n}\cdot P^{-1}\hat{\mathbf{J}}P,
\end{equation}
where $\mathbf{n}$ is the unit vector along the molecular field.
$\langle J_{\mathbf{n}}\rangle$ is the expectation value of the magnitude of $\hat{\mathbf{J}}$ polarized along $\mathbf{n}$ due to the molecular field.
$J_{\text{ex}}$ ($\sim 0.36$ K) was estimated by the de Gennes scaling from the $T_{\text{N}}$ of \ce{GdAuGe} \cite{kurumaji2023single}, and $\langle J_{\mathbf{n}}\rangle$ ($\sim 3.5$) was selected to be self-consistent with the angular momentum of diagonalized ground state ($\ket{GS}$ ($=\sum_{i=0\pm,1\pm}a_{i}\ket{i}$)) of the total Hamiltonian: $\langle J_{\mathbf{n}}\rangle=|\bra{GS}\hat{\mathbf{J}}\ket{GS}|$.

We set the orientation of the molecular field in various directions, and calculated the quadrupole moments for each orientation.
Five components of quadrupole moments are defined as
\begin{eqnarray}
Q_{x^2-y^2}&=&Q_0\bra{GS}(\hat{J}_x^2-\hat{J}_y^2)\ket{GS},\\
Q_{3z^2-r^2}&=&Q_0\bra{GS}\sqrt{3}(\hat{J}_z^2-\frac{J(J+1)}{3})\ket{GS},\\
Q_{yz}&=&Q_0\bra{GS}(\overline{\hat{J}_y\hat{J}_z})\ket{GS},\\
Q_{zx}&=&Q_0\bra{GS}(\overline{\hat{J}_z\hat{J}_x})\ket{GS},\\    
Q_{xy}&=&Q_0\bra{GS}(\overline{\hat{J}_x\hat{J}_y})\ket{GS},
\end{eqnarray}
where bars over symbols indicate the sum with respect to all possible permutations of the indices \cite{shiina1998interplay}.
$Q_0$ is defined by $-e\frac{\sqrt{3}}{2}\langle r^2\rangle g_9^{(2)}$ in Ref. \onlinecite{kusunose2008description}, where $e$ is the elementary charge, $\langle r^2 \rangle$ is the raidal average, and $g_9^{(2)} (=\frac{-2}{3^2\cdot 5\cdot 7})$ is a generalized Stevens' multiplicative factor for a Dy$^{3+}$ ion.
We note that we used the operators of the total angular momentum ($\hat{J}_i$) to calculate the quadrupole moment.
Namely, $Q_{ij}$ is directly related to the quadrupole moments in a Dy$^{3+}$ ion.
We may also be able to introduce effective quadrupole operators, $\hat{Q}^{\text{eff}}_{ij} (=\overline{\hat{S}_i\hat{S}_j}$, etc), by quadratic forms of effective spin operators $\hat{S}_i$ of $S_{\text{eff}}=3/2$ for the lowest quasi-quartet.
These would be useful to construct a low-energy effective model, while each $\hat{Q}_{ij}^{\text{eff}}$ is not directly proportional to $\hat{Q}_{ij}$, but can be related by a linear transformation such as $\hat{Q}^{\text{eff}}_{ij}=\sum_{m,n}c_{ijmn}\hat{Q}_{mn}$.
The parameters $c_{ijmn}$ are determined by the projection operator from $J=15/2$ space to the quartet space.

\section*{Data availability}
The datasets generated during and/or analysed during the current study are available in the Caltech Reserch Data Repository.

%


\begin{acknowledgements}
T.K. was supported by Ministry of Education Culture Sports Science and Technology (MEXT) Leading Initiative for Excellent Young Researchers (JPMXS0320200135) and Inamori Foundation.
This study was supported by Japan Society for the Promotion of Science (JSPS) KAKENHI Grant-in-Aid (No. 21K13874, 20J10988, 23K13068, 22K14010, JP19H05826, 19H01835).
The single-crystal x-ray diffraction was performed at SPring-8 with the approval of the Japan Synchrotron Radiation Institute (JASRI) (Proposal No. 2022A1751).
The inelastic neutron scattering experiment was performed at the Materials and Life Science Experimental Facility of the J-PARC under a program (Proposal No. 2024S09).
The resonant x-ray scattering experiment under a magnetic field was performed at PF under the approval of the Photon Factory Program Advisory Committee (Proposal No. 2022G551 and 2023G093).
This work was partly performed using the facilities of the Materials Design and Characterization Laboratory in the Institute for Solid State Physics, the University of Tokyo.
The authors thank L. Ye for fruitful discussion, A. Ikeda for generously allowing the use of optical sensing instrument (Hyperion si155, LUNA) for thermal expansion and magnetostriction measurements, and A. Kikkawa for supporting the synthesis of polycrystalline DyAuGe sample for the INS experiment.
\end{acknowledgements}

\section*{Contributions}
T.K. conceived the project, synthesized single crystals, measured magnetization, and analyzed the crystal electric field.
M.G. prepared and characterized polycrystalline DyAuGe for INS, and measured specific heat and magnetostriction in single crystals.
S.K. performed the synchrotron x-ray experiment at SPring-8.
K.I. and T.R.Y. performed the inelastic neutron scattering experiment at J-PARC.
T.K., M.G., H.S., H.N., and T.A. performed resonant x-ray experiment at Photon Factory.
T.K. wrote the manuscript, and all the authors read and commented on the manuscript.

\section*{Competing interests}
The authors declare no competing interests.


\end{document}